\documentclass[multphys,vecphys]{svmult}

\usepackage{makeidx}         % allows index generation
\usepackage{graphicx}        % standard LaTeX graphics tool
\usepackage{multicol}        % used for the two-column index
\usepackage[bottom]{footmisc}% places footnotes at page bottom
\makeindex             % used for the subject index
\begin{document}

\title*{Relativistic Iron lines at high redshifts}
\author{Andrea Comastri\inst{1} \and Marcella Brusa\inst{2} \and
 Roberto Gilli\inst{1}} 
\institute{INAF-Osservatorio Astronomico di Bologna, via Ranzani 1, I--40127
  Bologna, Italy
\texttt{andrea.comastri@oabo.inaf.it}, \texttt{roberto.gilli@oabo.inaf.it}
\and Max Planck Institut f\"ur Extraterrestrische Physik, Giessenbachstr. 1,
D--85748 Garching, Germany  \texttt{marcella@mpe.mpg.de}}
\maketitle

\begin{abstract}

The shape and the intensity of the 6.4 keV iron line
bring unique information on the geometrical and physical properties 
of the supermassive black hole and the surrounding accreting gas 
at the very center of Active Galactic Nuclei.
While there are convincing evidences of a 
relativistically broadened iron line in a few nearby bright objects,
their properties at larger distances are basically unknown.
We have searched for the presence of iron line by fully exploiting
{\it Chandra} observations in the deep fields. The line is clearly 
detected in the average spectra of about 250 sources stacked 
in several redshift bins over the range z=0.5-4.0. 
We discuss their average properties with particular enphasys 
on the presence and intensity of a broad component.

\end{abstract}

\section{Introduction}

Most of the accretion power, which makes Active Galactic Nuclei 
luminous hard X--ray sources, is released in the innermost regions 
around 
the central Supermassive Black Hole (SMBH), where the relativistic effects 
in the strong field limit (gravitational redshift and light bending)
significantly affect the emerging spectrum.
In particular the iron emission line at 6--7 keV is by far the most
significant feature in the AGN X--ray spectra and a detailed study
of its profile provides unique information about the gas properties 
and the nature of the spacetime in the proximity of the SMBH
(see \cite{com:fabmin} for an exhaustive review including the most recent 
observational results).
\par
A key issue concerns the shape of the line profile and in particular of the 
red wing, which is much broader if the emitting gas is accreting 
over a rapidly spinning SMBH. If this is the case,
the innermost stable orbit is much closer ($\simeq 1.24 r_g$ where 
$r_g = GM/c^2$) to the central BH  than for a non rotating BH ($\simeq6r_g$)
and the emitted photons suffer from stronger general relativistic effects 
which are in principle visible in the line profile provided that 
enough sensitivity is reached in the 2--10 keV band. 
\par 
The profile of the broad emission line in the nearby Seyfert galaxy
MCG--6--30--15 represents a textbook case. The presence of 
a relativistic line with a skewed profile extending down to
4 keV (and during some episodes even at lower energies 
\cite{com:iwa96}), was first unambiguously determined by ASCA \cite{com:tan95},
and then confirmed by essentially all the X--ray experiments 
(i.e. {\it BeppoSAX}, XMM--{\it Newton} and {\it Chandra}).
While the evidence for a line profile distorted by general relativistic
effects is quite solid in MCG--6--30--15, and tested against alternative 
solutions such as complex absorption \cite{com:young05}, 
the observational constraints 
on other nearby Seyfert galaxies are not such to rule out alternative 
possibilities. Moreover the search for broad iron lines
beyond the nearby Universe is strongly hampered by the sensitivity 
of the present instruments and only a few tentative positive results
were reported \cite{com:iron,com:min06}
\par 
Deep XMM--{\it Newton} and {\it Chandra} observations offer the 
possibility to search for iron line emission at much larger distances
($z >$0.5) at least in a statistical sense, by stacking the X--ray counts of
a large number of X--ray sources.

%%%%%%%%%%%%%%%%%%%%%%%%%%%%%%%%%%%%%%%%%%%
\begin{figure}
\centering
\includegraphics[height=15cm]{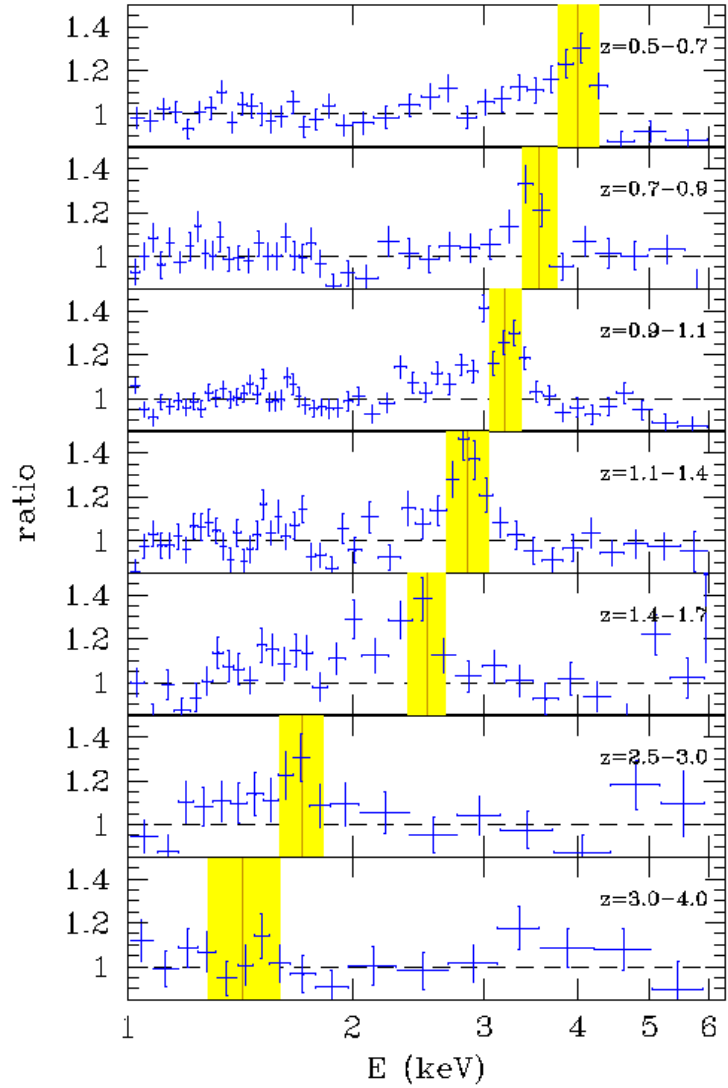}
\caption
{Residuals of a
simple power--law fit to the source spectra in seven different
redshift bins as labeled. The vertical line in each panel is at the
expected position for the redshifted 6.4 keV Fe $K\alpha$ line while
the shaded region encompasses the bin width reported in Table~1 and
defined as $\Delta$E = 6.4/(1+$z_{max}$) -- 6.4/(1+$z_{min}$) keV.}
\label{com:fig1}       % Give a unique label
\end{figure}
%%%%%%%%%%%%%%%%%%%%%%%%%%%%%%%%%%%%%%%%%%%

\section{Stacking in XMM--{\it Newton} and {\it Chandra} deep fields}

The individual spectra of 53 Type 1 AGN and 43 Type 2 AGN 
in the deep XMM--{\it Newton} pointing ($\sim$ 800 ksec) of the Lockman Hole
for which a spectroscopic redshift is available, 
were brought to rest--frame and then summed together \cite{com:alina}.
The residuals of a single power law fit to the stacked spectra 
clearly show a prominent iron line at 6.4 keV with a skewed profile
extending towards lower energies. A good fit to both (type 1 and type 2) 
line profiles is obtained with a disc model with an innermost orbit 
around $\sim$ 3 $r_g$, suggesting that, on average, most of the 
SMBH in distant AGN are spinning.
The above described approach maximizes the counting statistic in the 
energy band covered by the line, but at the same time looses, by definition,  
any information about the redshift dependence of line properties.
\par
A complementary approach \cite{com:bibi} has been pursued 
by stacking the X--ray counts, in the observed frame, 
of spectroscopically identified sources 
in the {\it Chandra} deep fields (CDFN \& CDFS) in seven redshift 
slices spanning the $z$=0.5--4 range. The choice of bin sizes
and distribution is driven by a trade--off between the number of
counts in each bin and the need to sample an observed
energy range narrow enough to detect the spectral feature, keeping at
the same time the instrumental response as uniform as possible.  
The sample includes 171 sources in the CDFN and 181 in the CDFS,
spanning the luminosity range L$_{2-10 keV}=10^{41}-10^{45}$ ergs s$^{-1}$. 
With the exception of the highest redshift bin, a significant excess
above a power--law continuum is present at the expected
energy (see Fig. \ref{com:fig1}). 

The residuals leftward of the iron line (which 
are more prominent in the bins with the highest number of sources 
and counting statistics $z$=0.5--0.7 and 0.9--1.1)
suggest the presence of a broad redshifted component, 
similar to that observed by XMM \cite{com:alina}.

\section{Searching for broad lines in deep fields} 

The stacking results described above, motivated by two 
different key scientific objectives, were obtained following two different 
approaches in the stacking procedure. 
The XMM--{\it Newton} stacked spectra were obtained by unfolding the 
instrumental response with a power law model. In this way it
is possible to shift and add the individual spectra and maximize
the S/N ratio around the iron line energy. The drawback is 
a model dependent parameterization of the underlying continuum.
%%%%%%%%%%%%%%%%%%%%%%%%%%%%%%%%%%%%%%%%%%%
\begin{figure}[bottom]
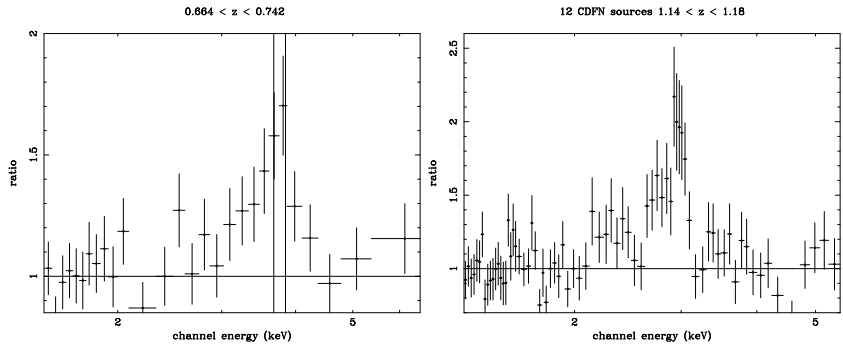

\centering
\includegraphics[width=4.45cm, angle=-90]{comastriF2a.ps}
\includegraphics[width=4.45cm, angle=-90]{comastriF2b.ps}
\caption
{{\it Left panel}: The residuals of a single power law fit to the  
stacked spectra of the 37 X--ray sources in the CDFS redshift slice 
0.664 $< z <$ 0.742. The vertical lines corresponds to the expected 
position of the 6.4 keV line at the two extremes of the interval. 
{{\it Right panel}:  
The fit and residuals of a single power law fit to the  
stacked spectra of the 12 objects in the CDFN redshift 
slice 1.14 $< z <$ 1.18. The expected position of the redshifted 6.4 keV
line is coincident with the peak in the residuals at $\sim$ 2.96 keV}}
\label{com:fig2}       % Give a unique label
\end{figure}
%%%%%%%%%%%%%%%%%%%%%%%%%%%%%%%%%%%%%%%%%%%
The alternative strategy \cite{com:bibi}, following a more conventional
approach, does not suffer of the model dependent unfolding of the 
instrumental response, but at the same time it is not well suited for a study 
of the line profile which is smeared in each redshift bin by 
the bin size itself.
A third possibility which combines the pros of the two above mentioned 
methods would be to stack a large number of source spectra  
within a redshift range
which is small enough that the energy (redshift) spread is negligible
or at least smaller than the instrumental energy resolution. 
\par
\begin{figure}[!top]
\centering
\includegraphics[width=4.65cm,angle=-90]{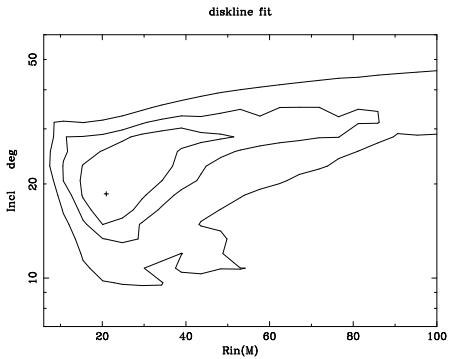}
\caption
{68, 90 and 99 \% confidence contours of the disk inner radius (units of 
$r_G$ and inclination angle (degrees)}
\label{com:fig3}       % Give a unique label
\end{figure}
\begin{figure}
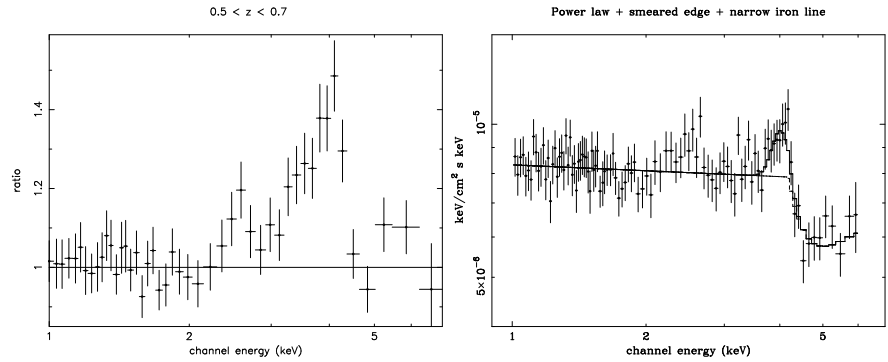

\centering
\includegraphics[width=4.65cm, angle=-90]{comastriF4a.ps}
\includegraphics[width=4.65cm, angle=-90]{comastriF4b.ps}
\caption
{{\it Left panel}: The residuals of a single power law fit to the  
$z$=0.5--0.7 bin adapted from \cite{com:bibi}.
{{\it Right panel}: The unfolded spectrum of a complex absorption model  fit 
plus a narrow gaussian emission line at 6.4 keV (rest--frame) 
to the same data set}}
\label{com:fig4}       % Give a unique label
\end{figure}

The presence of significant spikes in the redshift distribution 
observed in both CDFS \cite{com:ennis} and CDFN \cite{com:bar05}
has prompted us to further investigate such a possibility.
The residuals of a single power law fit to the  
stacked spectra of the 37 X--ray sources in the CDFS redshift slice 
(0.664 $< z <$ 0.742) and the 12 objects in the CDFN redshift 
slice (1.14 $< z <$ 1.18) are shown in Fig.~\ref{com:fig2}. In both cases, 
a broad wing is clearly visible redward of a narrow core corresponding 
to the redshifted energy of the 6.4 keV line.  
In both samples the line shape is well fitted by a relativistic disk 
line model. However, the accuracy in the determination of the 
inner disk radius is not good enough to constrain the SMBH 
spin (Fig.~\ref{com:fig3}). While this is not surprising given the available
counting statistics, the similarity of the very shape of the 
observed profile to what expected by a relativistic line is tantalizing.
Although the perspectives of the detection of spinning BH
at high redshift are promising a few words of caution appear to be 
appropriate.
It is well known that the line intensity and especially the 
line profile are dependent on an accurate modeling of the underlying continuum 
and in particular are sensitive to the presence of complex absorption 
\cite{com:fabmin}.
%%%%%%%%%%%%%%%%%%%%%%%%%%%%%%%%%%%%%%%%%%%
%%%%%%%%%%%%%%%%%%%%%%%%%%%%%%%%%%%%%%%%%%%
The latter possibility is especially relevant for the analysis of the 
stacked spectrum of the deep field sources which are mostly obscured AGN. 
The 2--7 keV continuum resulting from the 
superposition of absorbed spectra with different redshifts is modified
with respect to a single power law, by the absorption cut--off and most
important by the 7.1 keV iron edge. The net effect of the latter 
can be approximated with a smeared edge which effectively enhance the 
X--ray continuum just below the iron line emission.
The residuals of a single power law fit to the sources in the 
z=0.5--0.7 bin suggest the presence of an extended broad wing.
(Fig.~\ref{com:fig4}, left panel) 
which is almost completely accounted for by the
continuum predicted by X--ray background synthesis models \cite{com:gilli01}  
in the same redshift bin (Fig.~\ref{com:fig4}, right panel).  

%%%%%%%%%%%%%%%%%%%%%%%%%%%%%%%%%%%%%%%%%%%
%%%%%%%%%%%%%%%%%%%%%%%%%%%%%%%%%%%%%%%%%%%

\section{Conclusions}

The detection of spinning BH at high redshift is probably close 
to the capabilities of present instrumentation provided that deeper 
observations and/or larger samples of sources are collected and 
appropriately analyzed especially for what concern the 
modelling of the underlying continuum. 
Future missions with large collecting area 
will surely provide a step forward in such a direction.

%%%%%%%%%%%%%%%%%%%%%%%% referenc.tex %%%%%%%%%%%%%%%%%%%%%%%%%%%%%%
% sample references
% "physics"
%
% Use this file as a template for your own input.
%
%%%%%%%%%%%%%%%%%%%%%%%% Springer-Verlag %%%%%%%%%%%%%%%%%%%%%%%%%%

%
% BibTeX users please use
% \bibliographystyle{}
% \bibliography{}

%%%%%%%%%%%%%%%%%%%%%%%%%%%%%%%%%%%%%%%%%%%%%%%%%%%%%%%%%%%%%%%%%%%%%%  }

%%%%%%%%%%%%%%%%%%%%%%%%%%%%%%%%%%%%%%%%%%%%%%%%%%%%%%%%%%%%%%%%%%%%%%

\printindex
\end{document}